\documentclass[longauth]{aa}
\usepackage[utf8]{inputenc}
\usepackage{amsmath}
\usepackage{amsfonts}
\usepackage{amssymb}
\usepackage{color,soul}
\usepackage{longtable}
\usepackage{graphicx}

\begin{document}

%\begin{frontmatter}

\title{Tensile Strength of 67P/Churyumov-Gerasimenko Nucleus Material from Overhangs}
\author{
N.~Attree \inst{1} \and O.~Groussin \inst{1} \and L.~Jorda \inst{1} \and D.~N{\'e}bouy \inst{1}
\and N.~Thomas \inst{2}
\and Y.~Brouet \inst{2}
\and E.~K{\"u}hrt \inst{3} \and F.~Preusker \inst{3} \and F.~Scholten \inst{3} \and J.~Knollenberg \inst{3}
\and P.~Hartogh \inst{4}
\and H.~Sierks\inst{4}
\and C.~Barbieri\inst{5}
\and P.~Lamy\inst{1}
\and R.~Rodrigo\inst{6,7}
\and D.~Koschny\inst{6}
\and H.~Rickman\inst{9,10}
\and H.~U.~Keller\inst{3,11}
%\and J.~Agarwal\inst{2}
\and M.~F.~A'Hearn\inst{12}
%\and F.~Angrilli\inst{12}
\and A.-T.~Auger\inst{1}
\and M.~A.~Barucci\inst{13}
\and J.-L.~Bertaux\inst{14}
\and I.~Bertini\inst{5}
\and D.~Bodewits\inst{12}
\and S.~Boudreault\inst{4}
%\and D.~Bodewits\inst{11}
%\and C.~Capanna\inst{1}
\and G.~Cremonese\inst{19}
\and V.~Da Deppo\inst{17}
\and B.~Davidsson\inst{7}
\and S.~Debei\inst{16}
\and M.~De Cecco\inst{18}
\and J.~Deller\inst{4}
\and M.~R.~El-Maarry\inst{2}
%\and F.~Ferri\inst{15}
\and S.~Fornasier\inst{13}
\and M.~Fulle\inst{19}
%\and R. Gaskell\inst{4}
%\and L.~Giacomini\inst{15}
%\and P.~Gutierrez-Marques\inst{2}
\and P.~J.~Guti\'errez\inst{20}
\and C.~G\"uttler\inst{4}
%\and N.~Hoekzema\inst{2}
\and S.~Hviid\inst{3}
\and W.-H Ip\inst{21}
\and G.~Kovacs\inst{4}
\and J.~R.~Kramm\inst{4}
\and M.~K\"uppers\inst{22}
%\and F.~La~Forgia\inst{3}
\and L.~M.~Lara\inst{20}
\and M.~Lazzarin\inst{5}
%\and C.~Leyrat\inst{13}
\and J.~J.~Lopez Moreno\inst{20}
\and S.~Lowry\inst{23}
%\and S.~Magrin\inst{\bf ?}
\and S.~Marchi\inst{24}
\and F.~Marzari\inst{5}
%\and H.~Michalik\inst{26}
%\and R.~Moissl\inst{23}
\and S.~Mottola\inst{4}
\and G.~Naletto\inst{5, 17, 15}
\and N.~Oklay\inst{3}
\and M.~Pajola\inst{27}
%\and M.~Pertile\inst{15,16}
%\and F.~Preusker\inst{3}
%\and L.~Sabau\inst{28}
%\and F.~Scholten\inst{3}
%\and C.~Snodgrass\inst{28}
\and I.~Toth\inst{28}
\and C.~Tubiana\inst{4}
\and J.-B.~Vincent\inst{3}
\and X.~Shi\inst{4}
%\and M.~Zaccariotto\inst{15,16}
%\and P.~Wenzel\inst{6}
       }
       
  \institute{\tiny Aix Marseille Univ, CNRS, LAM, Laboratoire d'Astrophysique de Marseille, Marseille, France, \email{Nicholas.Attree@lam.fr} %1
        \and
        Physikalisches Institut, Sidlerstr. 5, University of Bern, CH-3012 Bern, Switzerland%2
        \and
        Institute of Planetary Research, DLR, Rutherfordstrasse 2, 12489, Berlin, Germany%3
        \and
        Max-Planck-Institut f\"ur Sonnensystemforschung, 37077 G\"ottingen, Germany%4
        \and
        Department of Physics and Astronomy, Padova University, Vicolo dell'Osservatorio 3, 35122, Padova, Italy%3
        \and
        Centro de Astrobiologia (INTA-CSIC), 28691 Villanueva de la Canada, Madrid, Spain %4
        \and
        International Space Science Institute, Hallerstrasse 6, CH-3012 Bern, Switzerland %5
        \and
        Scientific Support Office, European Space Agency, 2201, Noordwijk, The Netherlands%6
        \and
        Department of Physics and Astronomy, Uppsala University, Box 516, 75120, Uppsala, Sweden%7
        \and
        PAS Space Research Center, Bartycka 18A, PL-00716 Warszawa, Poland%8
        \and
        Institute for Geophysics and Extraterrestrial Physics, TU Braunschweig, 38106, Germany%10
        \and
        Department of Astronomy, University of Maryland, College Park, MD, 20742-2421, USA%11
        \and
        LESIA, Obs. de Paris, CNRS, Univ Paris 06, Univ. Paris-Diderot, 5 place J. Janssen, 92195 Meudon, France %13
        \and
        LATMOS, CNRS/UVSQ/IPSL, 11 boulevard d'Alembert, 78280, Guyancourt, France%14
        \and
        Centro di Ateneo di Studi ed Attivit{\`a} Spaziali, "Giuseppe Colombo" (CISAS), University of Padova, via Venezia 15, 35131 Padova, Italy%15
        \and
        Department of Industrial Engineering, University of Padova, 35131 Padova, Italy%16
        \and
        CNR-IFN UOS Padova LUXOR, via Trasea 7, 35131 Padova, Italy%17
        \and
        UNITN, Universit di Trento, via Mesiano, 77, 38100 Trento, Italy%18
        \and
        INAF - Osservatorio Astronomico, Via Tiepolo 11, 34143, Trieste, Italy%20
        \and
        Instituto de Astrofisica de Andaluc\'ia (CSIC), Glorieta de la Astronom\'ia s/n, 18008 Granada, Spain%21
        \and
        Institute for Space Science, Nat. Central Univ., 300 Chung Da Rd., 32054, Chung-Li, Taiwan%22
        \and
        Operations Department, European Space Astronomy Centre/ESA, P.O. Box 78, 28691 Villanueva de la Canada, Madrid, Spain%23
        \and
        Centre for Astrophysics and Planetary Science, School of Physical Sciences (SEPnet), The University of Kent, Canterbury, CT2 7NH, UK
        \and
        Southwest Research Institute, 1050 Walnut St., Boulder, CO 80302, USA%25
        \and
        INAF, Osservatorio Astronomico di Padova, 35122 Padova, Italy%26         %\and
        %Institut f\"ur Datentechnik und Kommunikationsnetze der TU Braunschweig, Hans-Sommer-Str. 66, 38106 Braunschweig, Germany
        \and
        University of Padova, Department of Information Engineering, Via Gradenigo 6/B, 35131 Padova, Italy%27
        \and
        NASA Ames Research Center, Moffett Field, CA 94035, USA %28
%         \and
%         Instituto Nacional de Tecnica Aeroespacial, 28850 Torrejon de Ardoz (Madrid), Spain
        %\and
        %Planetary and Space Sciences, Department of Physical Sciences, Open University, Milton Keynes MK7 6AA, UK%
        \and
        MTA CSFK Konkoly Observatory, H1121 Budapest, Konkoly Thege M. ut 15-17%29
  }

\abstract{
We directly measure twenty overhanging cliffs on the surface of comet 67P/Churyumov-Gerasimenko extracted from the latest shape model and estimate the minimum tensile strengths needed to support them against collapse under the comet's gravity. We find extremely low strengths of around one Pa or less (one to five Pa, when scaled to a metre length). The presence of eroded material at the base of most overhangs, as well as the observed collapse of two features and implied previous collapse of another, suggests that they are prone to failure and that true material strengths are close to these lower limits (although we only consider static stresses and not dynamic stress from, for example, cometary activity). Thus, a tensile strength of a few pascals is a good approximation for the tensile strength of 67P's nucleus material, which is in agreement with previous work. We find no particular trends in overhang properties with size, over the $\sim10-100$ m range studied here, or location on the nucleus. There are no obvious differences, in terms of strength, height or evidence of collapse, between the populations of overhangs on the two cometary lobes, suggesting that 67P is relatively homogenous in terms of tensile strength. Low material strengths are supportive of cometary formation as a primordial rubble pile or by collisional fragmentation of a small (tens of km) body.}


\maketitle

\section{Introduction}
Material strength is an important parameter in constraining the formation and evolution of comets and in explaining their morphological diversity. Low strengths support the conclusion that comets are primordial rubble piles, accreted gently (collision velocities of $\sim1$ m~s$^{-1}$ to tens of m~s$^{-1}$) in the early solar system, as opposed to remnants from higher velocity collisions (at hundreds of m~s$^{-1}$ to km~s$^{-1}$) which would undergo impact compaction, leaving them with higher strengths \citep{Davidsson16}. The strength of cliffs and overhangs against collapse also directly relates to cometary activity as cliff collapses are an important source of jets and outbursts \citep{Vincent16, Pajola2017}.

The tensile strength of cometary nucleus material has been estimated for a number of comets and using a number of different methods including: observations of comet breakups, from rotation (see for example \citealp{Davidsson01}) or close Solar \citep{Klinger, Steckloff} or Jupiter \citep{Asphaug96} encounters; laboratory experiments \citep{Blum06, Blum14, Bar-Nun07}; computer modelling \citep{Greenberg95, Biele09}; and, in the case of 67P/Churyumov-Gerasimenko (hereafter 67P), observations of cliffs, overhangs and boulders.  As summarised by \cite{Groussin15}, strength estimates vary over several orders of magnitude for cometary material but are generally low: in the range of pascals to tens of pascals for tensile strength ($\sigma_{T}$) at the metre scale, and larger for shear and compressive strengths. The discussion of scale is important because consolidated material generally shows decreasing strength at larger scales, following a power law proportional to $d^{-q}$, with length scale $d$ and an exponent of $q\sim0.6$ for water ice \citep{Petrovic2003}.

For 67P, the high quality of OSIRIS (Optical, Spectroscopic, and Infrared Remote Imaging System) imaging allowed the examination of a variety of features at the metre and decametre scale and, from this, \cite{Groussin15} measured the tensile strength of an overhang and a collapsed feature, estimating $\sigma_{T}=1-3$ Pa and $<150$ Pa at the $5-30$ m scale. When scaled to 1 m, this gives $\sigma_{T}=8-39$ Pa and $<1150$ Pa for these features, which are located in the Imhotep and Maftet regions, on the head and body of the comet, respectively (see \citealp{Thomas15} and \citealp{ElMaarry15} for a detailed description of 67P's regions). \cite{Vincent17} also measured the heights of cliffs and derived $1-2$ Pa strengths at the decametre scale across 67P's surface in order for the cliffs not to collapse. 

The goal of this work is to perform a more comprehensive survey of overhanging cliffs across comet 67P's surface, in order to quantify material strengths and its homogeneity or heterogeneity. In Section \ref{method} we describe the method developed to identify and measure overhanging cliffs. The results are presented in Section \ref{results} and discussed in Section \ref{discussion} and conclusions drawn in Section \ref{conclusion}.

\section{Method}
\label{method}

%\begin{figure}
%\begin{center}
%\includegraphics[width=16.45cm]{slopes.eps}
%\caption{}
%\label{slopes}
%\end{center}
%\end{figure}

We measure overhang properties using a digital terrain model (DTM or shape model) of comet 67P, constructed using the Stereo Photogrammetry (SPG) technique from Rosetta/OSIRIS images. \cite{Groussin15} showed that the alternative SPC technique can underestimate the number of high-slope facets compared to SPG, making SPG a more appropriate technique for studying sharp topography. We use the latest DLR shape model (SPG-SHAP7, \citealp{Preusker17}) with 22 million vertices, with a typical spacing of 1-2 m and orientation uncertainty of 2-5$^{\circ}$. For computational reasons, we use a decimated version (2 million facets, $\sim7$ m lateral vertex spacing) of the full shape model. Gravitational vectors are calculated for each facet following the method of \cite{Jorda2012}, which includes centrifugal forces, and are then compared to the facet normal vectors to produce a map of gravitational slope. Locally flat regions have a gravitational slope of $0^{\circ}$, while vertical cliffs are at $90^{\circ}$ and anything larger is an overhang. There are $15,358$ facets ($0.77\%$ of the total number) with slopes greater than $90^{\circ}$ and approximately 5000 greater than $100^{\circ}$. A number of these are artefacts of the reconstruction but many are real overhanging features, visible in OSIRIS images.

Due to the complicated shape of the nucleus, and the presence of artefacts in the reconstructed shape model, it was deemed impossible to automatically detect and characterise every overhanging feature. Instead we identify, by eye, a number of features and investigate them in detail. The features are selected by picking the most obvious large groupings of overhanging facets on the shape model, from all over the nucleus. Selecting the largest (actually `deepest') overhangs places the strongest constraints on strength. We examine the spatial distribution of selected features to ensure good coverage in Section \ref{results}.

\begin{figure}
\resizebox{\hsize}{!}{\includegraphics{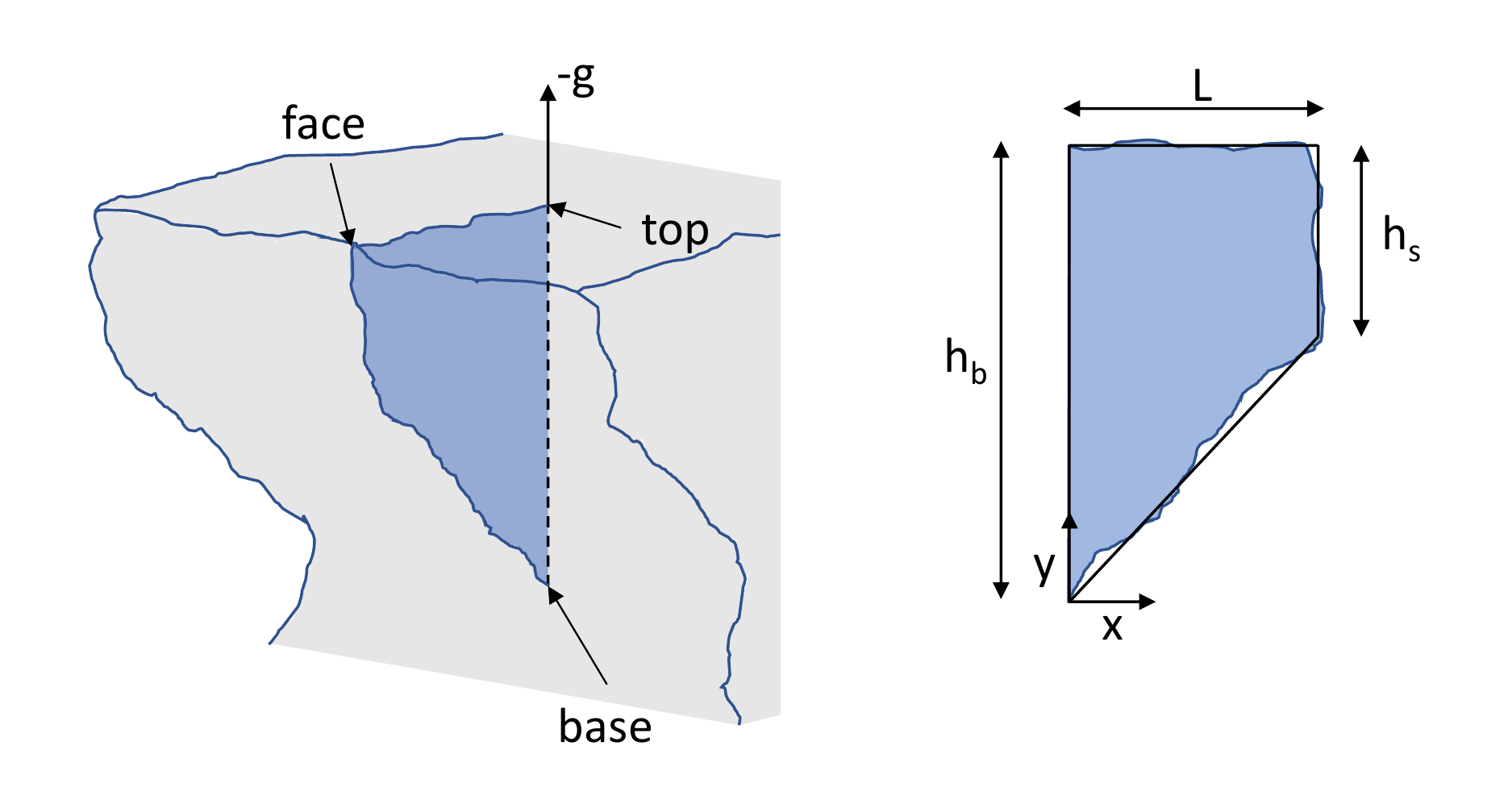}}
\caption{Overhang measurement method using the coordinates of three facets, at the base, top and face, of the overhang to define a plane and finding its intersection with the shape model. Aligning the profile with the local gravity vector, g, then allows the various overhang parameters to be measured, as in \cite{Tokashiki2010}.}
\label{overhang}
\end{figure}

For each selected feature, a local DTM is extracted from the full model, and a vertical profile of the overhang is computed in the following way. First, the negatives of the gravitational vectors are plotted on each high-slope facet. These point `up', ie.~away from the centre of gravity, and, in an overhang region, will penetrate inside the shape model before exiting it again at the top of the overhang (see Figure \ref{overhang}). The two facets where the gravitational vector enters and exits the nucleus define the coordinates of the `base' and `top' of the overhang. Selecting a third point as a facet on the `face' of the overhang, perpendicular to the gravity vector, defines a plane, the intersection of which with the shape model can be calculated. The final step is to project the coordinates of the intersection points onto the plane itself and then rotate the data to align them with the local gravity vector. This vector will vary across the overhang but the differences over such short distances ($\sim100$ m) are negligible, and the `base' gravity vector is used for the whole profile here. A profile of the shape model along the chosen vector is then recovered.

From such a profile, an estimate of the strength of the overhang can be calculated using the equations of a cantilevered beam. Following \cite{Tokashiki2010} and using the coordinate frame centred on the overhang base (as shown in Fig.~\ref{overhang}), the maximum stress from bending is found at $x=0$ and decreases along the length. The material tensile strength must be at least as large as this maximum stress to prevent immediate failure and collapse. For a beam of unit thickness and length $L$ and height $h$, this can be expressed as
\begin{equation}
\sigma_{T} > \frac{6M}{h^{2}},
\label{stress}
\end{equation}
where $M=\int^{L}_{0} x\rho gh\ dx$ is the total bending moment acting on the cantilever from its own unit weight of: unit density, $\rho=537.8$ kg m$^{-3}$ \citep{Preusker17}, times the magnitude of the local gravity, $g$. For simple beam shapes, this can be integrated directly. In a rectangular beam, for example, $h$ is constant and the bending moment evaluates to $M=h\rho g L^{2}/2$ while stress becomes
\begin{equation}
\sigma_{T} > \frac{3\rho g L^{2}}{h}.
\label{simplestress}
\end{equation}
For a trapezium shaped beam, as shown in Fig.~\ref{overhang}, $h$ varies along the length but the integral can still be evaluated, so that this time the stress becomes
\begin{equation}
\sigma_{T} > \frac{6\rho g L^{2}}{h_{b}} \left(\frac{1}{2} - \left(\frac{1-\alpha}{3}\right)\right),
\label{trapezestress}
\end{equation}
where $\alpha=h_{s}/h_{b}$ is the ratio of the height at the far-end to the base height. These two equations are useful approximations for many overhangs and were used in the previous works \citep{Groussin15, Tokashiki2010}. In our case, however, we have the full overhang profile from the shape model and can therefore integrate this directly, without having to use one of these approximations. We do this numerically by, first, interpolating the profile shape to a regularly spaced series of $x$ positions (separated by $dx$), measuring $h$ at each of these and then making the sum $\sum^{N}_{0} x\rho gh\ dx$, where $N=L/dx$.

As can be seen from Eqs.~\ref{simplestress} and \ref{trapezestress}, the overhang length is the most important parameter for constraining the material strength, and this remains true for the numerical integration. Therefore, for each of our overhangs we select the intersection plane to compute the profile along the deepest part of the overhang, by choosing the `base' facet with a gravity vector penetrating most deeply into the shape model (maximum L). In most cases, this is obvious from visual inspection and in cases with several similar facets, similar strength estimates will be derived.

There is an uncertainty in the position of all coordinates measured on the shape mode, which can be conservatively estimated as the average radius of a facet. For our two million facet model this is $\sim1.6$ m. In addition, the uncertainty in the orientation of each facet has been estimated as $\Delta\theta\approx\pm5^{\circ}$ \citep{Jorda2012} and in practice this can dominate over the positional uncertainty. We therefore estimate the uncertainties in overhang proportions by rotating each profile by $\pm\Delta\theta$, computing the integral and measuring $h$, and then combining this uncertainty with the position error. Uncertainties in tensile strength are then derived using the standard error propagation formulae.

\section{Results}
\label{results}

Table \ref{resultstab} shows the results for the 20 overhanging features analysed. Overhang heights are between $\sim10$ and 100 m and depths are generally only $\sim10$ m (Fig.~\ref{hL}), with a single feature (number 2) having a larger, almost 40 m, depth. This feature, part of a cliff in the Babi region facing Hapi, is visually the largest overhang on the shape model but is otherwise unremarkable. 

\renewcommand{\arraystretch}{1.2}  % ROW SEPARATION HERE: 0.5 CAN BE CHANGED.
\begin{table*}
\begin{center}
\begin{tabular}{cccccccccc}
No. & Latitude ($^{\circ}$) & Longitude ($^{\circ}$) & \multicolumn{2}{c}{h (m)} & \multicolumn{2}{c}{L (m)} & \multicolumn{2}{c}{$\sigma_{T}$ (Pa)} \\
\hline
1 & 16.21 & -67.47 & 109 & $^{+1.8}_{-2.4}$ &  13 & $^{+8.8}_{-7.2}$ & 0.16 & $\pm0.164$ \\ 2 & 21.29 & 60.84 & 113 & $^{+6.0}_{-3.6}$ &  38 & $^{+8.0}_{-8.1}$ & 1.64 & $\pm1.025$ \\ 3 & -56.01 & -101.77 & 100 & $^{+11.1}_{-9.9}$ &  15 & $^{+6.7}_{-6.8}$ & 0.26 & $\pm0.234$ \\ 4 & 40.12 & 23.93 & 109 & $^{+7.5}_{-15.2}$ &  11 & $^{+7.6}_{-5.1}$ & 0.15 & $\pm0.152$ \\ 5 & 41.94 & 140.53 &  67 & $^{+4.4}_{-6.9}$ &  11 & $^{+4.4}_{-4.3}$ & 0.17 & $\pm0.149$ \\ 6 & -56.22 & 58.67 & 106 & $^{+6.2}_{-99.3}$ &   6 & $^{+8.1}_{-5.8}$ & 0.04 & $\pm0.072$ \\ 7 & 32.32 & 161.04 &  41 & $^{+3.7}_{-11.6}$ &   6 & $^{+3.1}_{-3.1}$ & 0.07 & $\pm0.070$ \\ 8 & -49.00 & 88.69 &  32 & $^{+2.1}_{-2.8}$ &   5 & $^{+2.7}_{-2.3}$ & 0.12 & $\pm0.103$ \\ 9 & -4.69 & -7.72 &  33 & $^{+22.2}_{-14.0}$ &   2 & $^{+2.2}_{-2.0}$ & 0.01 & $\pm0.020$ \\ 10 & 3.33 & -122.62 &  37 & $^{+1.8}_{-1.6}$ &  12 & $^{+2.7}_{-2.7}$ & 0.58 & $\pm0.366$ \\ 11 & 10.90 & -129.46 &  61 & $^{+2.8}_{-3.6}$ &  11 & $^{+4.1}_{-4.1}$ & 0.24 & $\pm0.193$ \\ 12 & 20.43 & 150.23 &  95 & $^{+8.2}_{-47.1}$ &   8 & $^{+4.5}_{-2.4}$ & 0.08 & $\pm0.090$ \\ 13 & 41.99 & 7.99 &  32 & $^{+1.6}_{-8.1}$ &   7 & $^{+2.7}_{-2.7}$ & 0.18 & $\pm0.137$ \\ 14 & -20.69 & 17.02 &  43 & $^{+2.3}_{-11.4}$ &   7 & $^{+3.2}_{-3.2}$ & 0.13 & $\pm0.113$ \\ 15 & 68.89 & -159.29 &  65 & $^{+1.6}_{-2.7}$ &  11 & $^{+4.7}_{-4.8}$ & 0.25 & $\pm0.208$ \\ 16 & -19.06 & 99.47 &  42 & $^{+2.8}_{-2.8}$ &   9 & $^{+2.9}_{-2.8}$ & 0.27 & $\pm0.202$ \\ 17 & -33.69 & 117.56 &  47 & $^{+3.8}_{-4.5}$ &   9 & $^{+2.7}_{-2.8}$ & 0.19 & $\pm0.148$ \\ 18 & 39.06 & -125.97 &  48 & $^{+2.3}_{-2.7}$ &   8 & $^{+3.1}_{-2.8}$ & 0.22 & $\pm0.168$ \\ 19 & 11.62 & 107.20 &  56 & $^{+2.8}_{-5.8}$ &  10 & $^{+3.1}_{-3.1}$ & 0.23 & $\pm0.185$ \\ 20 & -21.81 & -28.84 &   9 & $^{+1.6}_{-1.6}$ &   6 & $^{+1.7}_{-1.7}$ & 0.77 & $\pm0.414$ \\ 
\end{tabular}
\label{resultstab}
\caption{Locations (in the Cheops frame) and properties of each measured overhang. $h$ and $L$ are the heights and depths, as directly measured from the profiles, while $\sigma_{T}$ is derived by numerically integrating the profile shape with Eqn.~\ref{stress}.}
\end{center}
\end{table*}

\begin{figure}
\resizebox{\hsize}{!}{\includegraphics{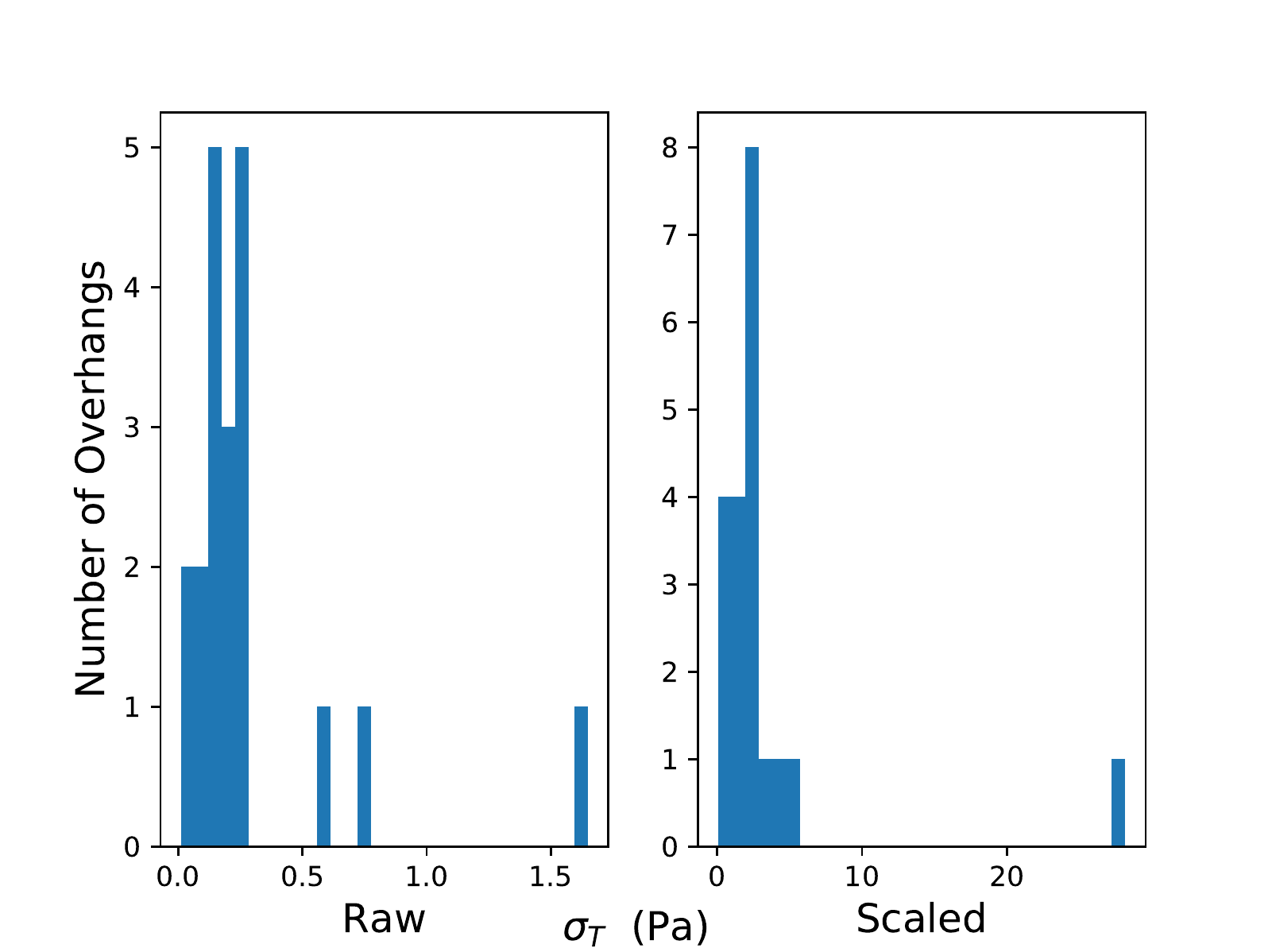}}
\caption{Lower limit of tensile strengths for the measured overhangs. On the right the data have been scaled from the feature length scale ($h$) to the equivalent strength at the metre scale.}
\label{sigma}
\end{figure}

The calculated overhang strengths are very low, with a mean of 0.3 Pa and an average uncertainty of $\pm 0.22$ Pa. Figure \ref{sigma} shows the distribution in measured strengths and those scaled from the feature length-scale, $h$, to 1 m, using the power law for ice (the scaling law for ice is used, despite the large dust content, as a first estimate and for ease of comparison with previous studies). The scaled strengths all lie between zero and 5 Pa, apart from the single large outlier of feature 2. Nonetheless, the uncertainty in sigma for this feature could easily bring its value in line with the others. We detect no particular relation between strength, scaled or raw, and overhang height, suggesting a uniform strength over the range of $\sim10-100$ m.

\begin{figure*}
\includegraphics[width=17cm]{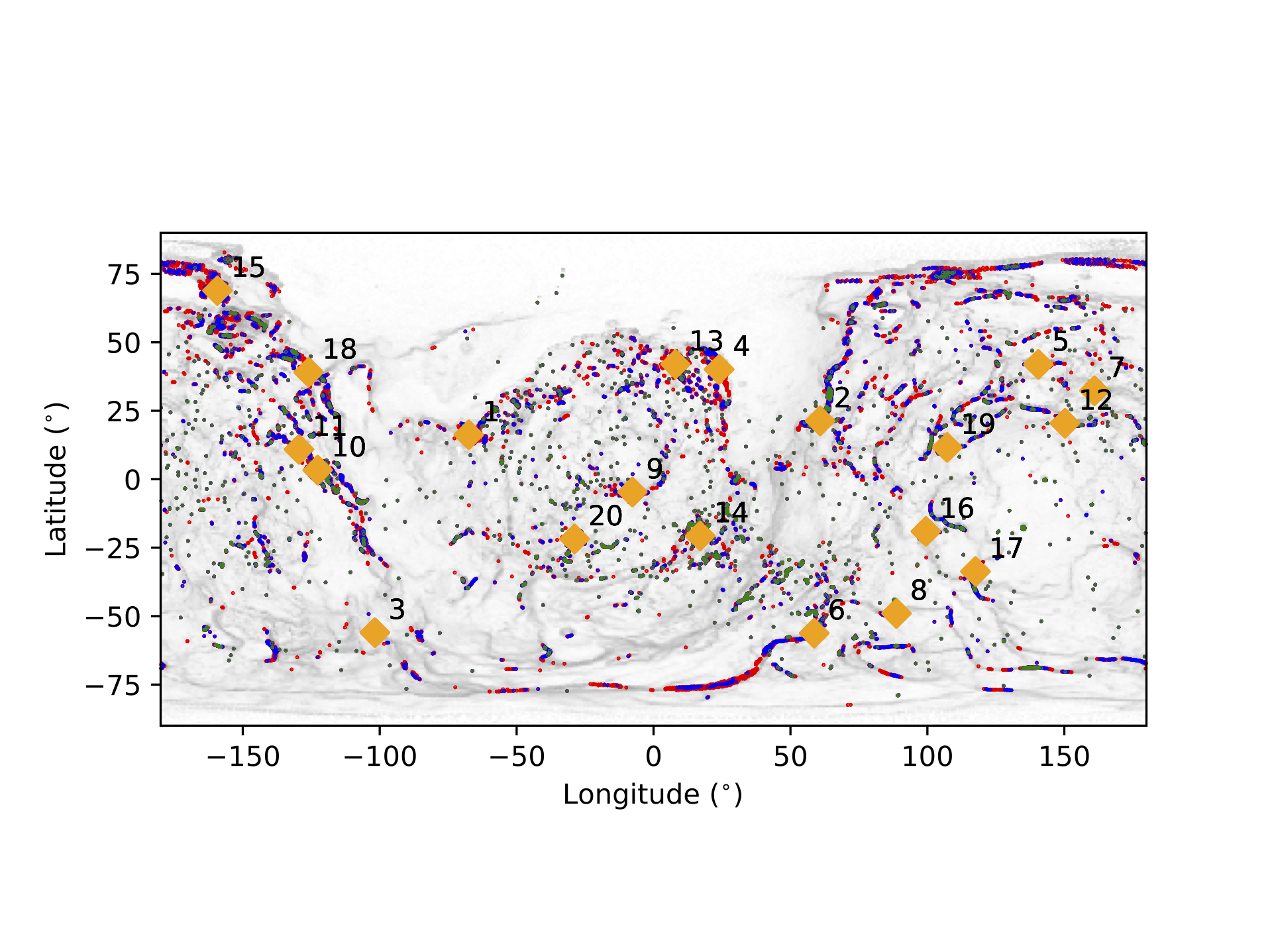}
\caption{Location of the measured overhangs (orange diamonds). Also shown are high-slope facets on the shape model with the colour scheme: green $\ge100^{\circ}$, $100^{\circ}>$ blue $\ge90^{\circ}$ and $90^{\circ}>$ red $\ge85^{\circ}$.}
\label{latlon}
\end{figure*}

\begin{figure*}
\includegraphics[width=17cm]{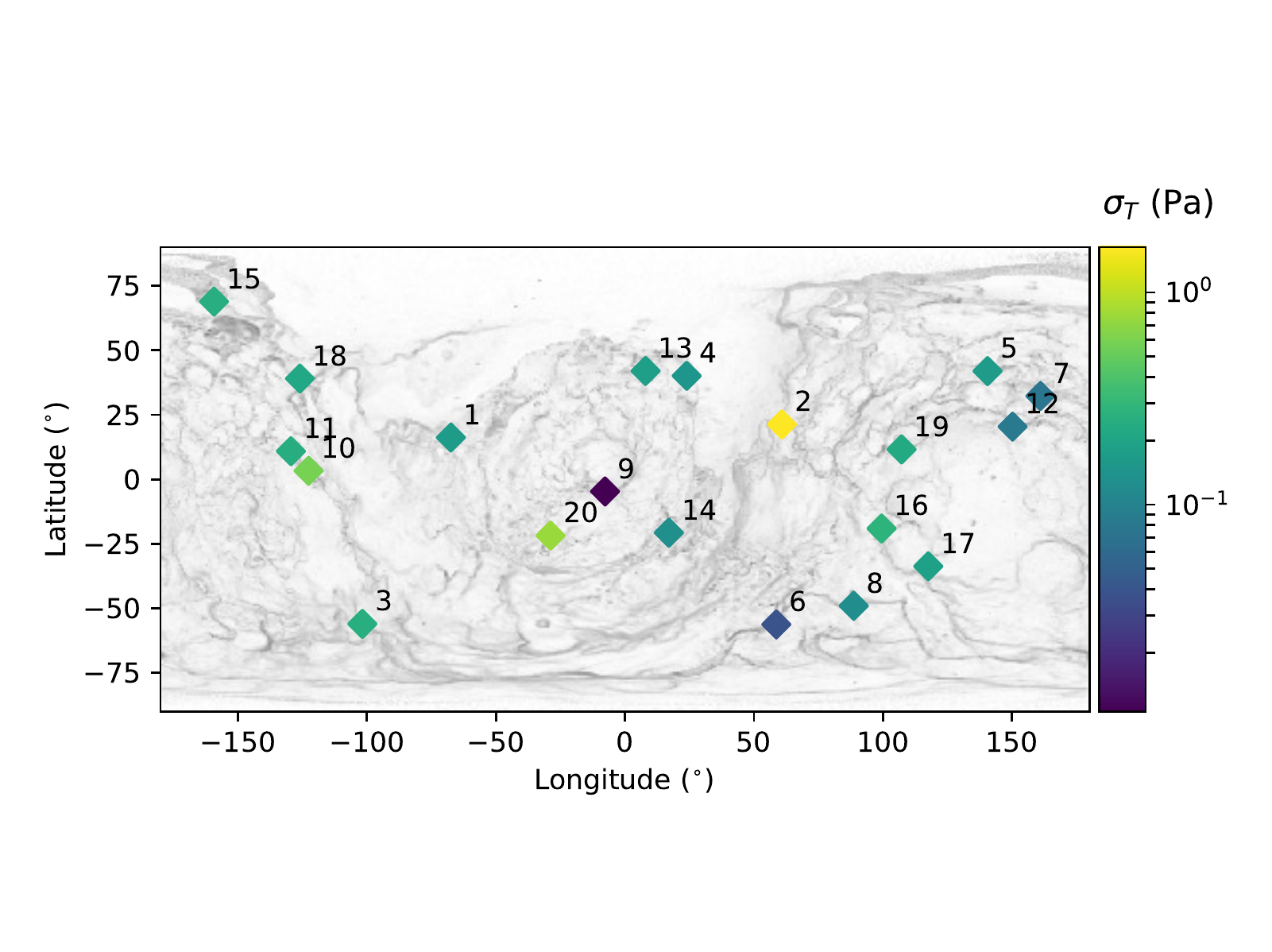}
\caption{Location of the measured overhangs. The colour scheme shows the (unscaled) tensile strength on a log scale.}
\label{latlon_c}
\end{figure*}

\begin{figure*}
\includegraphics[width=17cm]{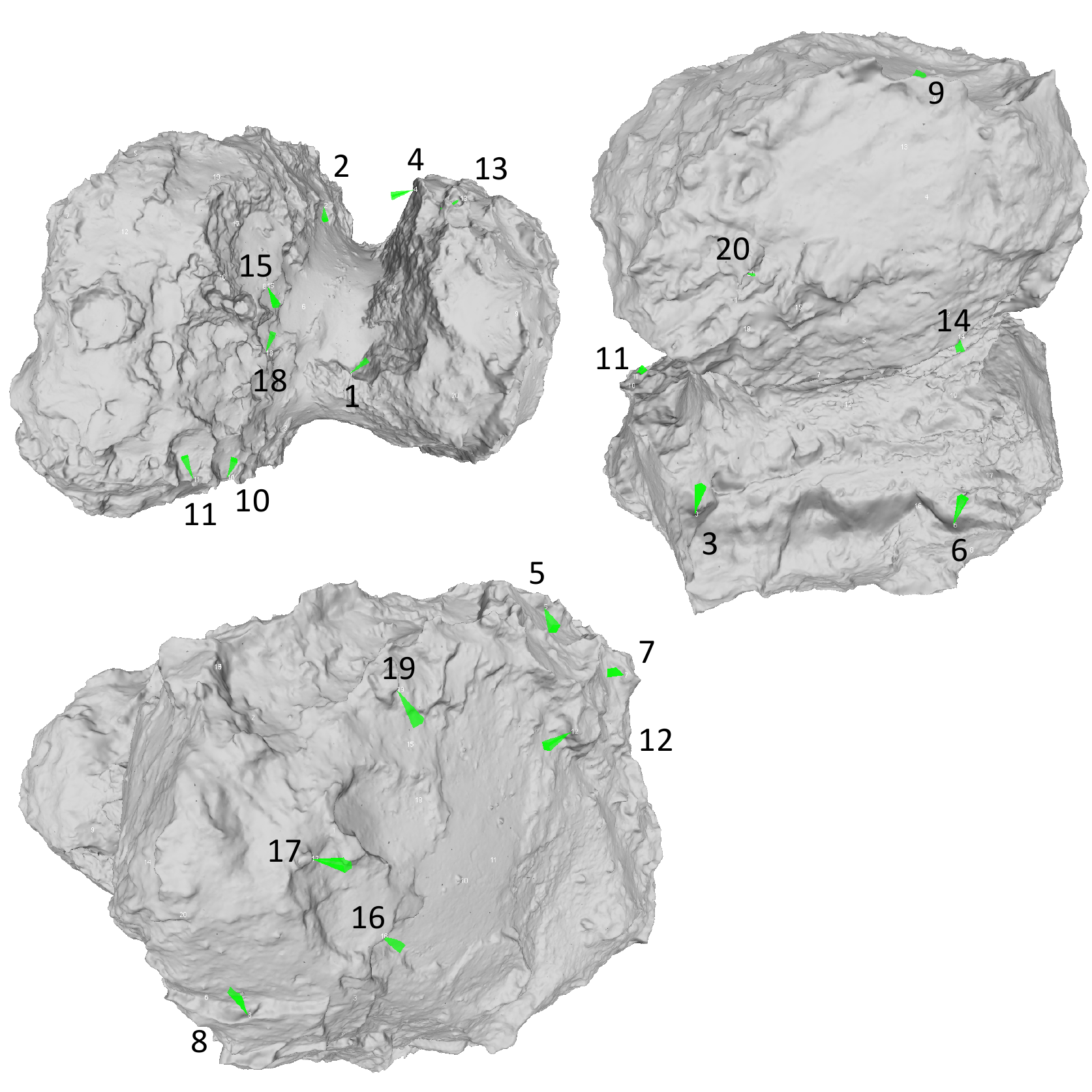}
\caption{Location of the measured overhangs on the shape model.}
\label{locations}
\end{figure*}

Figures \ref{latlon}, \ref{latlon_c} and \ref{locations} show the locations of each measured feature on the cometary surface, along with all the high-slope facets. High-slopes are typically clustered into curvilinear chains of cliffs. The only large area lacking high-slopes (other than artefacts) is the smooth, dusty terrain of Hapi. Our selected overhangs are well distributed across the surface, with features seen in both hemispheres, both cometary lobes and in a variety of cometary regions. No particular trends are noted between overhang properties and location, and there are no significant differences between the average strengths on each lobe, suggesting a uniformity in the capacity of 67P's material to support overhangs.

\begin{figure*}
\centering
\includegraphics[width=17cm]{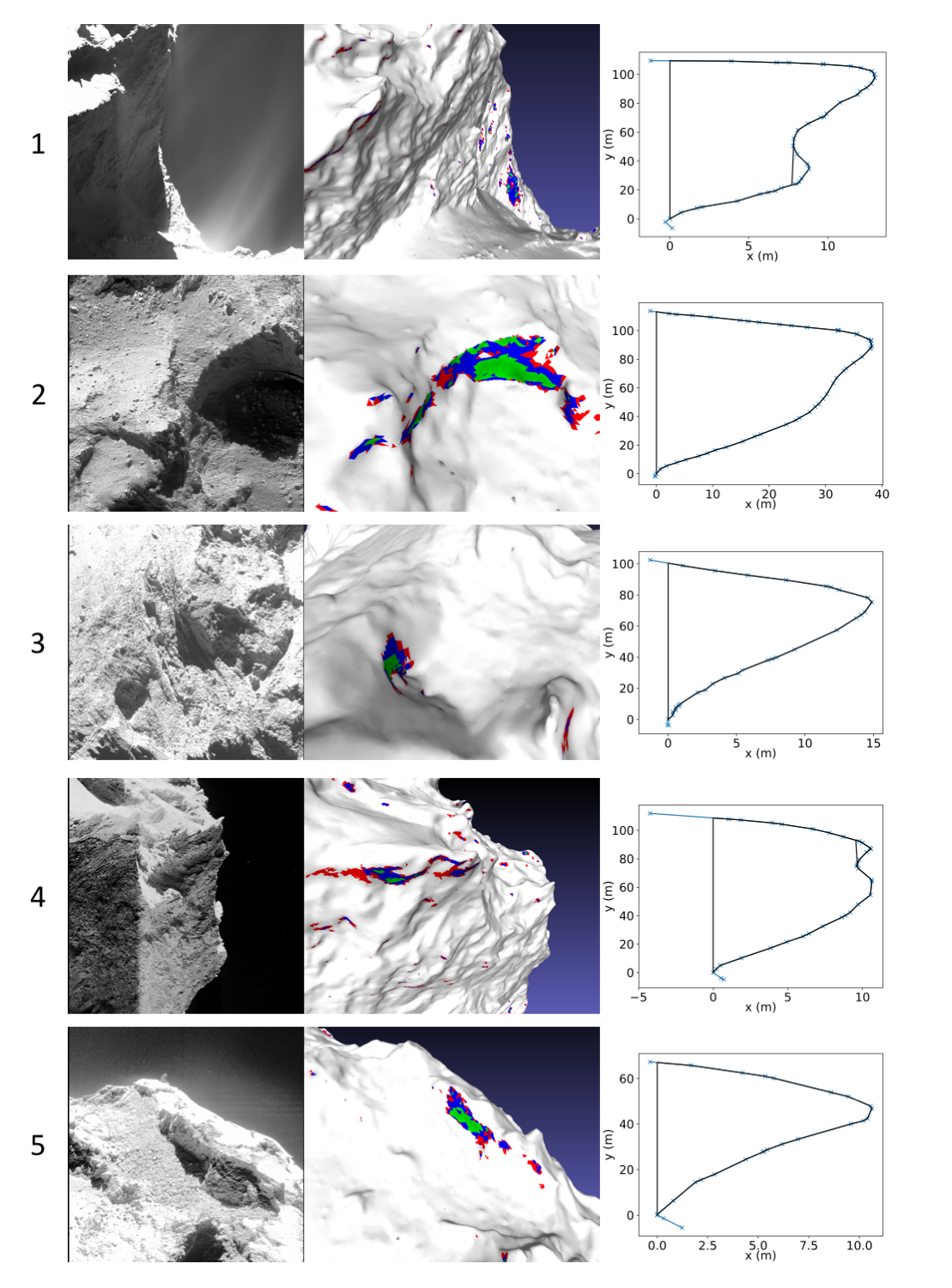}
\caption{Overhangs 1-5. On the left are representative, contrast enhanced, OSIRIS images: NAC\_2014-12-01T21.02.45, NAC\_2016-05-13T06.42.54, NAC\_2016-07-16T06.46.11, NAC\_2014-10-01T06.49.53 and NAC\_2014-10-01T04.36.23, respectively. A visualisation of the overhang on the shape model is shown in the middle, with facets colour-coded by slope as above (green $\ge100^{\circ}$, $100^{\circ}>$ blue $\ge90^{\circ}$ and $90^{\circ}>$ red $\ge85^{\circ}$). On the right, the shape model profile along an intersection through the overhang is shown in blue and our interpolated area for integration in black.}
\label{1}
\end{figure*}

\begin{figure*}
\centering
\includegraphics[width=17cm]{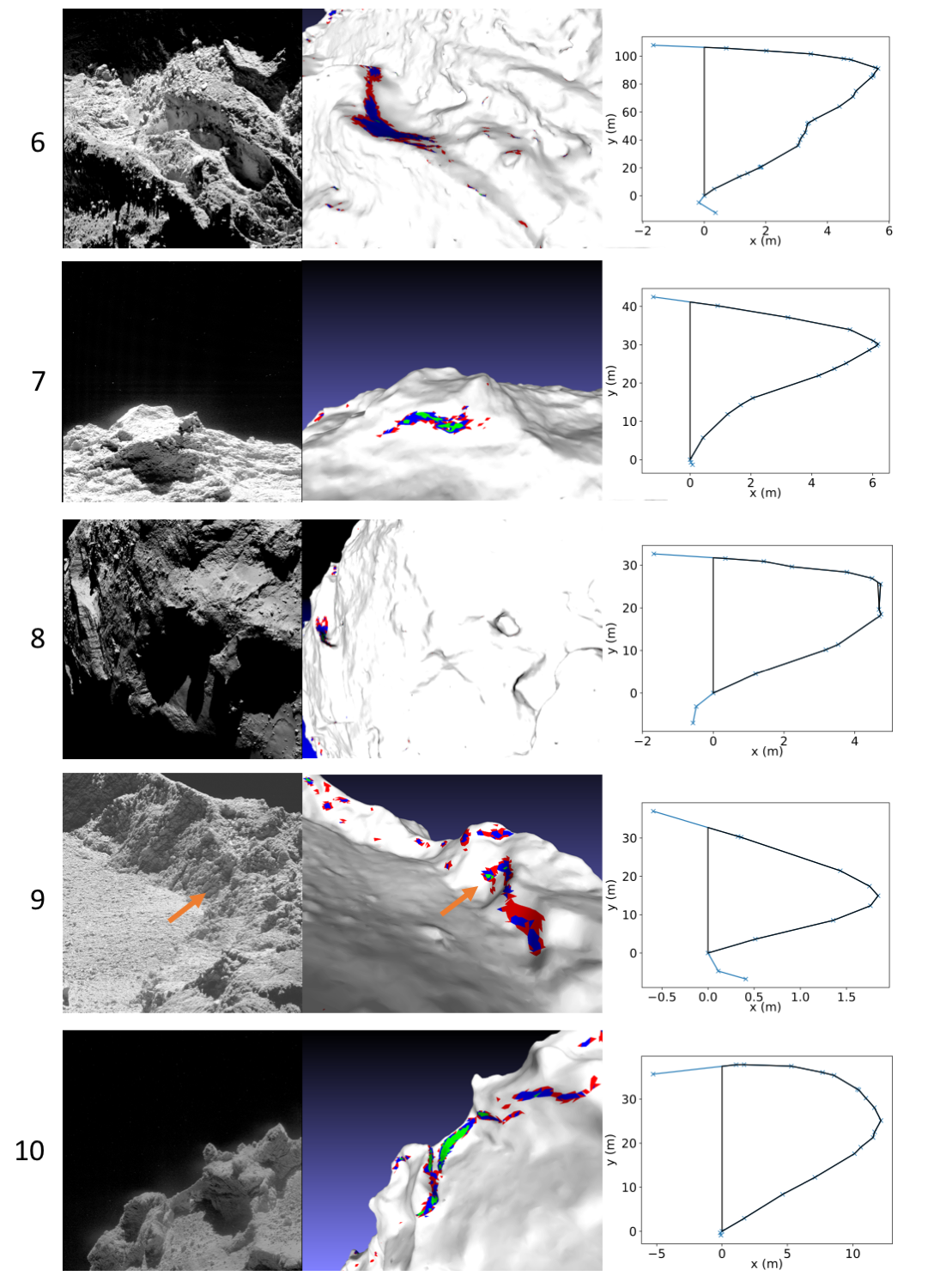}
\caption{Overhangs 6-10. Same as Fig.~\ref{1} with images NAC\_2016-07-09T03.28.54, NAC\_2014-10-05T16.55.16, NAC\_2016-06-07T21.07.00, NAC\_2016-05-20T08.01.06 and NAC\_2016-03-13T16.53.39, respectively.}
\label{2}
\end{figure*}

\begin{figure*}
\centering
\includegraphics[width=17cm]{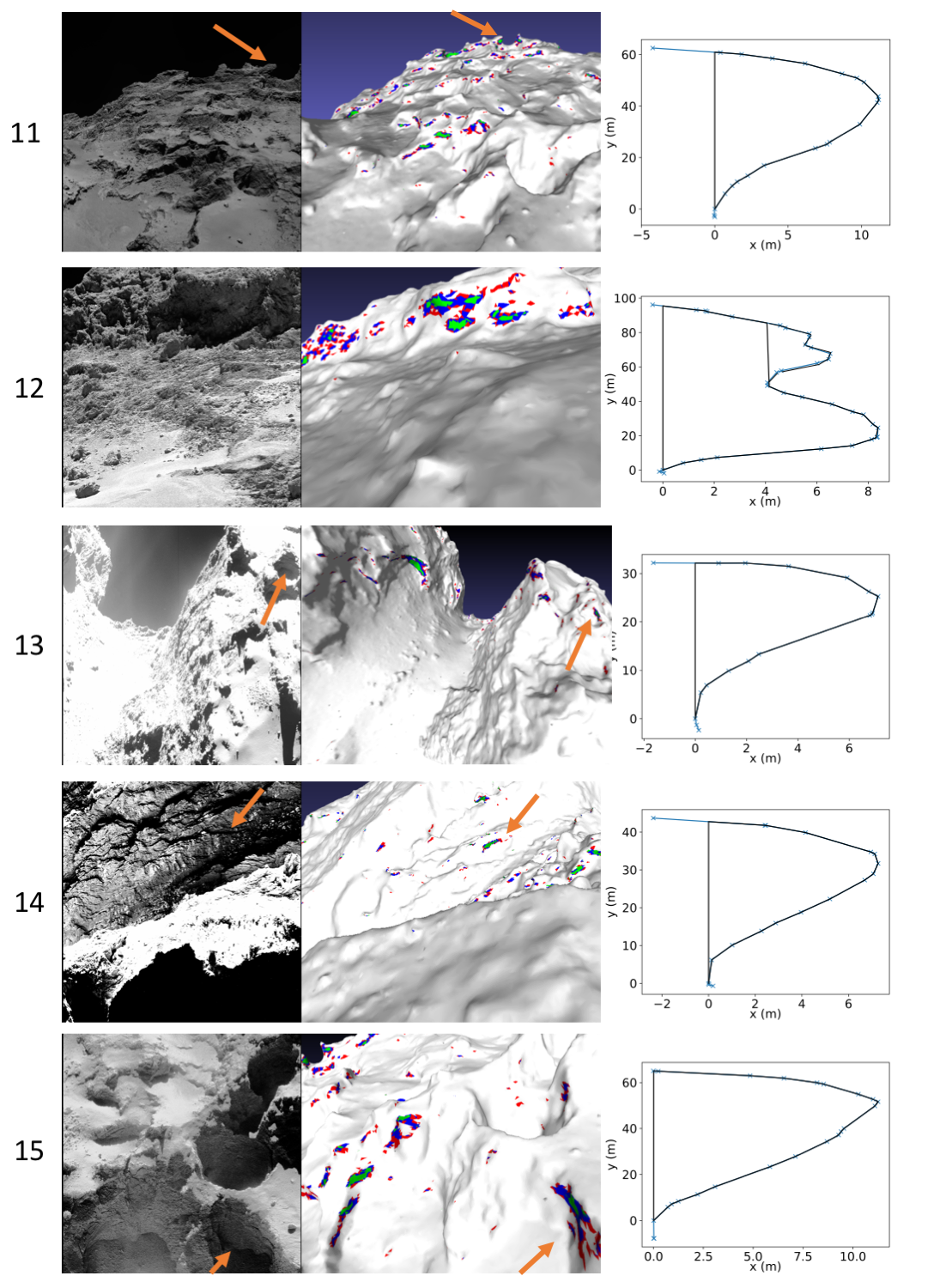}
\caption{Overhangs 11-15. Same as Fig.~\ref{1} with images NAC\_2014-09-22T01.37.07, NAC\_2016-07-09T21.44.48, NAC\_2014-09-23T09.42.48, NAC\_2016-04-23T18.12.52 and NAC\_2014-10-02T00.26.22, respectively.}
\label{3}
\end{figure*}

\begin{figure*}
\centering
\includegraphics[width=17cm]{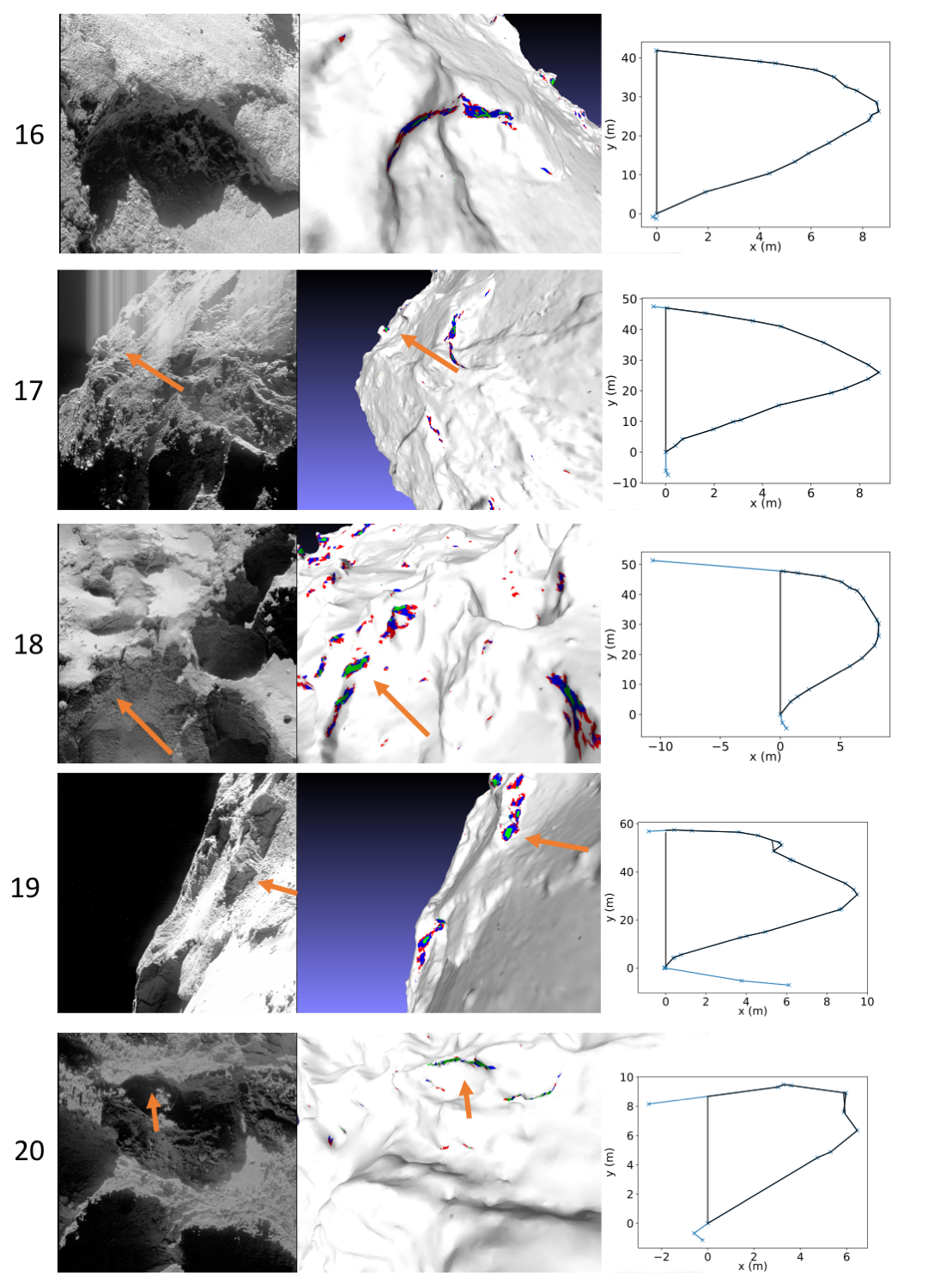}
\caption{Overhangs 16-20. Same as Fig.~\ref{1} with images NAC\_2016-03-19T23.04.57, NAC\_2015-01-16T01.44.08, NAC\_2014-10-02T00.26.22, NAC\_2016-04-29T15.55.45 and NAC\_2016-03-19T16.38.38, respectively.}
\label{4}
\end{figure*}

The features have a variety of morphologies, as can be seen in their profiles and accompanying OSIRIS images in Figs.~\ref{1} - \ref{4}. They range from large, shallow cliffs to protruding blocks and the lips of pit walls. Most have a shallow trapezium or triangular shape, with a number having curving top surfaces, and are not obvious `caves', but rather are steep cliffs which lean slightly `outwards' towards their tops, producing an overhanging region. Several features are actually double or triple overhangs (features 1, 4, 16 and 19 in Figs.~\ref{1} - \ref{4}), and for these we split the integration region before calculating the moment of each area and manually summing them together. Nearly all the overhangs examined here show some evidence of collapse or erosion, such as boulders and debris fields at their base. The exceptions are features 1, 4, and 17, where the possible debris are some distance away due to their positions at the top of tall cliffs, and 3 and 20, where their presence is slightly ambiguous.

Of particular interest are features 5, 12 and 15. Feature 12 is the same as the failed section already measured in \cite{Groussin15}. The overhang in Ash near feature 5 was seen to suffer a collapse between May and December 2015 (\citealp{ElMaarry2017} supplementary Fig.~2). We examine a local high-resolution DTM of the area (Preusker, F., personal communication) but were unfortunately unable to produce a profile of the collapsed region due to technical issues. Measuring directly from the DTM in a 3D model viewer, however, we estimate the overhanging segment to have $L\approx12$ m and $h\approx31$ m, giving $\sigma_{T}\approx0.4$ Pa with the rectangular approximation of Eqn.~\ref{simplestress}, in-line with the other overhangs measured here. Similarly, feature 15, located in Seth and named Aswan, was seen to collapse in July 2015 (triggering an outburst, \citealp{Pajola2017}). They measure the overhanging section as $65\times12$ m, again consistent with the $63\times11$ m, $\sigma=0.33$ Pa found here.

\section{Discussion}
\label{discussion}

\subsection{Are overhangs representative of bulk nucleus strength?}
The strengths estimated above are lower limits; the material must have at least this tensile strength, in order not to immediately collapse under its own weight, but could be much stronger. The presence of collapses, however, implies that overhangs may be close to failure and that these estimates may indeed be close to the actual material strength. 

Conversely, \cite{Vincent17} and \cite{Pajola2017} argue that cliff heights are not controlled by intrinsic material strength but by external erosion, in which case observed features would not be limited by gravity and tensile strengths could be larger. Many of the features examined here have clearly fractured cliff walls (e.g.~2, 6, 9, 10, 13, 15, 16, 18), some reminiscent of thermal contraction crack polygons \citep{Auger15, Auger17}, which may imply additional processes, such as thermal stresses and sublimation, weakening them before failure occurs. Additionally, material strength can vary on a local scale. For example, results from the MUPUS and SESAME experiments on the Philae lander \citep{Spohn15, Knapmeyer17} as well as theoretical \citep{Kossacki15} and laboratory work \citep{Grun93, Kochan89} suggest a hard, ice bonded layer with much greater strength within $\sim$ metres of the surface. The SESAME results in particular suggest a layer at depths of $10-50$ cm with a tensile strength on the order of MPa \citep{Knapmeyer17}.

In order to consider these possibilities, we compare the dimensions of the measured overhangs with the depths of a hard layer, and to those of temperatures relevant for thermal processing, using a comet thermal model described in \cite{Attree17}. Our model takes into account a spherical nucleus (orientated according to its pole with RA = 69.57\,deg, DEC = 64.01\,deg (J2000) and with a rotational period of P = 12.40\,hr; \citealp{Jorda2016}), solar insulation, and heat conductivity \citep{GroussinLamy}. We compute the temperature on the surface and inside the nucleus over one complete revolution, taking into account the diurnal and seasonal changes in insolation with heliocentric distance. To ensure convergence, we use a time step of 12.4\,s and ran the thermal model over 5 complete revolutions. We then compute the maximum temperature over an orbital period, experienced at each depth interval (for 2000 depth intervals of a thickness of a fifth of a diurnal skin depth each) for three different values of thermal inertia: I = 10, 50 and 250 J\,m$^{-2}$\,K$^{-1}$\,s$^{-1/2}$.

\begin{figure}
\resizebox{\hsize}{!}{\includegraphics{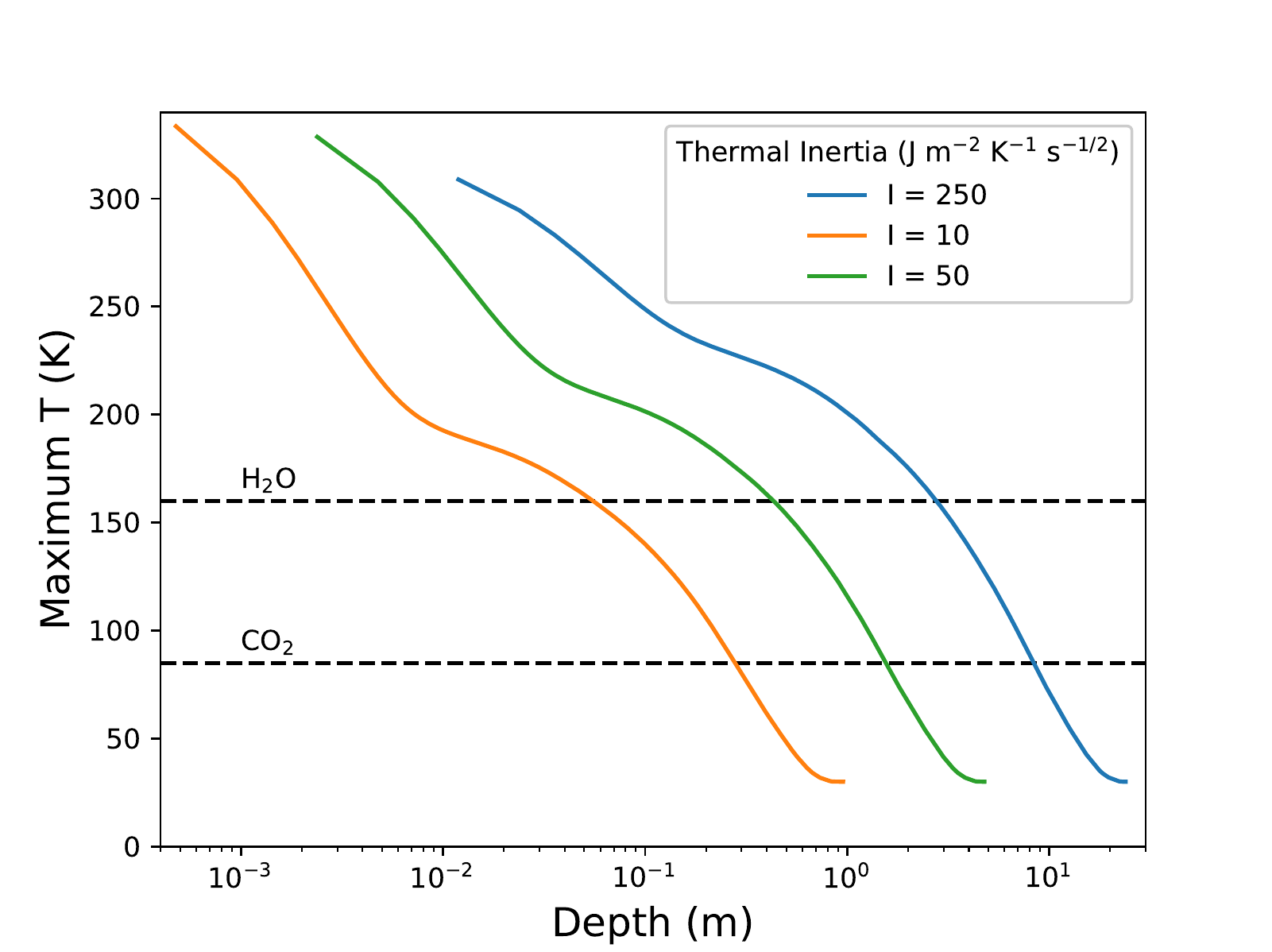}}
\caption{Maximum temperature reached with depth at the equator of 67P for three different values of thermal inertia. Horizontal lines indicate conservative estimates for the temperature where sublimation is negligible (compared to the erosion rate at perihelion) for water ice (160 K) and CO$_{2}$ (85 K).}
\label{Tprofiles}
\end{figure}

Figure \ref{Tprofiles} shows the resulting thermal profiles at the equator. The horizontal lines show conservative estimates for the temperatures at which sublimation is negligible compared to the erosion rate at perihelion (one thousandth its value) for water ice (160 K) and CO$_{2}$ (85 K). Depths of 2.8 m, for water ice, and 8.4 m, for CO$_{2}$, therefore define the limits to which we would expect sublimation to affect material properties. Our overhangs are typically $\sim10$ m deep and, with one exception (feature 9), all have $L>5$ m. Assuming H$_{2}$O is the dominant volatile component, we therefore expect our strength measurements to be probing material which has not been thermally processed. Further, since pressure inside the comet is on the order of tens to hundreds of Pa \citep{Groussin15}, thermal or compressional processing should not affect material below these depths and we expect the measured tensile strength to be a good approximation for that of the bulk nucleus.

A hard, ice bonded layer will increase the average strength of near surface material. The fact that we measure such low strengths, however, implies that the hard layer is localised and does not contribute significantly to the average strength of cometary material at depth. This is consistent with an estimated hard layer thickness of $0.1-0.5$ m, compared to the $\sim5-10$ m deep overhangs. Therefore, we conclude this section by reinforcing the idea that the tensile strengths measured above ($\sim0-5$ Pa when scaled to metre lengths) are indeed representative of bulk cometary tensile strengths at the decametre scale.

\subsection{Dynamic stresses}
Thus far our analysis has only focused on the strength of overhangs to resist the static stress of their own weight, however dynamic stresses should also be present on the comet. These might include stresses induced by rotational changes \citep{Stubbe16, Hirabayashi}, as well as seismic events, such as cometary activity/outbursts and impacts (e.g.~\citealp{ThomasRobinson}).

Impact features are rare on 67P's surface, and with only a few, small ($\sim10$s of metres) craters detected during the duration of the Rosetta mission \citep{ElMaary2017EPSC}, seem unlikely to provide a comet-wide mechanism for overhang collapse. Activity related stresses, on the other hand, are very likely to have occurred during the comet's approach of the Sun, but quantifying them remains difficult due to the still poorly understood activity mechanisms. Seismic energy, released from a particular source of cometary activity, should fall off with the square of the distance times some attenuation factor \citep{ThomasRobinson}, which may be large due to the comet's fractured and porous nature. As a first order estimate, the stress in a weak elastic wave is $\sigma=\rm{Uv\rho}$ \citep{Melosh} which, with a wave velocity of U $>80$ ms$^{-1}$ (as measured by experiments on Philae; \citealp{Knapmeyer17}) and assuming a particle velocity of v $=1$ cm~s$^{-1}$, will be at least $430$ Pa, greater than the strength estimates here. Continuous activity is not localised to particular regions, however, and even transient outburst features are found all-over the nucleus, suggesting it will hard to correlate the locations of activity with collapsed overhangs (the exception being where collapses \textit{trigger} outbursts, such as in the Aswan case above).

Due to the reaction force on the nucleus, activity can alter cometary orbit and rotation, and several studies have investigated the stresses induced by such changes in rotation pattern on 67P's complex shape. Both \cite{Stubbe16} and \cite{Hirabayashi} found large stresses of up to several hundred Pa centred on the neck region. These dynamic stresses indeed exceed static stress and would lead to failure if the material has the strengths measured here. Our measured overhangs (both those with and without collapse features) have no apparent correlation with the neck, however. Additionally, the presence of collapses in features 5 and 12, located on the big lobe, far form the neck, argues for static-stress induced failure. A full analysis of the correlation between activity- and rotation-driven stress patterns and the location of overhangs could put further constraints on the material strength but is beyond the scope of this paper. Here we must limit our conclusion to the following: that static stress analysis provides lower limits to cometary material strength, but which may indeed be close to the real values due to the presence of overhangs which appear to have collapsed under their own weight.

\section{Conclusion}
\label{conclusion}

We examine 20 overhanging cliffs, measuring their vertical profiles  using the up-to-date SPG SHAP7 shape model (Preusker et al. 2017) of comet 67P. From this we derive lower limits for the material's tensile strength, in order to support such overhangs against gravity. Overhangs are generally shallow (most have depths $\sim10$ m) and so the resulting tensile strengths are very small; $\sigma_{T}\sim1$ Pa or less at the decametre scale and $\sim0-5$ Pa when scaled to metre lengths (except for one outlier at $\sim28$ Pa, but with relatively large uncertainties). Nevertheless, the presence of eroded material at the base of most overhangs, and the observed collapse of two features and implied previous collapse of another, suggests that they are near to failure. Thus, $\sigma_{T}$ of a few pascals is a good estimate for the tensile strength of 67P's nucleus material, although further analysis of dynamic stresses, such as those caused by cometary activity and rotation changes, is warranted. Thermal modelling shows little material alteration at relevant depths in the subsurface, suggesting this value to be a reasonable approximation for bulk strengths at depth. This is in good agreement with previous estimates (as can be seen in the summary table in \citealp{Groussin15}) from modelling \citep{Greenberg95, Biele09}, laboratory experiments \citep{Bar-Nun07, Blum14} and some observations \citep{Asphaug96}, including cliff heights \citep{Vincent17}. Other observations, such as from the breakup of sungrazing and rotating comets \citep{Klinger, Davidsson01, Steckloff}, suggest somewhat higher values of tens of Pa to $\sim100$ Pa. The \cite{Groussin15} overhang results are slightly higher than those presented here because of their approximation of overhang shapes as rectangular, compared to the shape model profiles used here. 

We find no particular trends in overhang properties with size, over the $\sim10-100$ m range studied here, or location on the nucleus. There are no obvious differences, in terms of strength, height or evidence of collapse, between the populations of overhangs on the two cometary lobes, suggesting that 67P is relatively homogenous in terms of tensile strength.

Such a low and homogeneous strength has implications for the formation and evolution of 67P. In terms of evolution, low strengths mean that material is easily eroded by sublimation, gas pressure and thermal fracturing, and is vulnerable to collapse under its own gravity. Collapses naturally explain the retreating cliffs, with debris fields and fallen boulders at their feet, seen across the comet (see for example \citealp{Pajola15, ElMaarry2017, Pajola2017}), as well as the presence of the overhangs themselves. These may form from the partial collapse of sections of cliff following preexisting weaknesses or which are further weakened by thermal fracturing (see model of erosion of \citealp{Attree17}). Cometary outgassing activity has been linked to such collapses \citep{Pajola2017} and to active cliff faces in general \citep{Vincent16}, demonstrating that cliffs and overhangs are important areas of erosion on cometary surfaces.

Low bulk strengths support the conclusions of \cite{Davidsson16} that 67P represents a primordial rubble pile, directly accreted from the proto-solar nebular by hierarchical aggregation or streaming instabilities, or is a collisional fragment from a small (tens of km) body. A fragment, or rubble pile of fragments, from the disruption of a larger ($\sim1000$ km) body would inherit some of that body's properties, such as  higher density and strength material from impact compaction and/or thermal processing and differentiation. Low strength is more consistent with early formation at low collision velocities \citep{SkorovBlum12} in a dynamically cold disk, whilst a homogeneity between 67P's two lobes could imply a similar formation mechanism for both.

\begin{acknowledgements}
This project has received funding from the European Union's Horizon 2020 research and innovation programme under grant agreement no 686709. This work was supported by the Swiss State Secretariat for Education, Research and Innovation (SERI) under contract number 16.0008-2. The opinions expressed and arguments employed herein do not necessarily reflect the official view of the Swiss Government.

OSIRIS was built by a consortium of the Max-Planck-Institut f{\"u}r Sonnensystemforschung, G\"ottingen, Germany; the CISAS University of Padova, Italy; the Laboratoire d'Astrophysique de Marseille, France; the Instituto de Astrofisica de Andalucia, CSIC, Granada, Spain; the Research and Scientific Support Department of the ESA, Noordwijk, Netherlands; the Instituto Nacional de T{\'e}cnica Aeroespacial, Madrid, Spain; the Universidad Polit{\'e}chnica de Madrid, Spain; the Department of Physics and Astronomy of Uppsala University, Sweden; and the Institut f{\"u}r Datentechnik und Kommunikationsnetze der Technischen Universit{\"a}t Braunschweig, Germany. The support of the national funding agencies of Germany (DLR), France (CNES), Italy (ASI), Spain (MEC), Sweden (SNSB), and the ESA Technical Directorate is gratefully acknowledged. We thank the Rosetta Science Operations Centre and the Rosetta Mission Operations Centre for the successful rendezvous with comet 67P/Churyumov-Gerasimenko.
\end{acknowledgements}

\bibliographystyle{aa}
\bibliography{../../../papers/Bibliography}

\end{document}